# Large-scale Complex IT Systems

Ian Sommerville, Dave Cliff, Radu Calinescu, Justin Keen, Tim Kelly, Marta Kwiatkowska, John McDermid and Richard Paige

## 1. Introduction

On the afternoon of 6th May 2010, the US equity markets underwent an extraordinary upheaval. In about 10 minutes, the Dow Jones Industrial Average dropped by over 600 points representing the disappearance of around 800bn dollars of market value. In the course of this sudden downturn, the share-prices of several blue-chip multinational companies went crazy – shares in companies that had been a few tens of dollars plummeted to $0.01 in some instances, and rocketed to values over $100,000.00 in others.

As suddenly as this market downturn occurred, it reversed; over the next few minutes most of the loss was recovered. Share prices returned to levels within a few percentage points of the values they had held before the crash. This 'Flash Crash' sparked a major enquiry into its causes by the CFTC (Commodity Futures Trading Commission) and the SEC (Securities Exchange Commission).

Various theories were discussed in the five months that it took to produce the final report on the events of May $6^{th}$ (CFTC&SEC, 2010). Many speculated on the role of high-frequency trading (HFT) by investment banks and hedge funds, where algorithmic trading systems (algos) buy and sell blocks of financial instruments on incredibly short timescales, often holding a position for a second or less.

When the final report on the Flash Crash was finally published, it stated that the trigger-event for the crash was a single block-sale of $4.1bn worth of futures contracts, executed with uncommon urgency on behalf of a fund-management company. It was argued that the consequences of that trigger event interacting with algos rapidly buying and selling shares rippled out to cause the system-level failures.

The Flash Crash is an example of the kind of large-scale system failure that can arise as a consequence of software actions. The 'failure' was not caused by software bugs - rather, the interactions of independently-managed software systems created conditions that were unforeseen by any of the owners and developers of the trading systems. This led to a failure in the broader, socio-technical markets in which the algorithmic trading systems are used.

Our economy and society is becoming increasingly dependent on complex IT systems that are created by integrating and orchestrating independently managed software systems. We argue here that the incredible increase in scale and complexity in such systems means that we need new software engineering techniques that can help us cope with the inherent complexity in these systems. Without these, failures like the Flash Crash will become increasingly common and may have large-scale societal effects. In this article, we explain that there are principled reasons why current software engineering cannot scale and we propose a research and education agenda to help us address the problems of large-scale complex, IT systems engineering.

## 2. Coalitions of systems

The key factor that characterises large, complex IT systems is that these systems are assembled from other existing and new systems, which are independently controlled and managed. Current software engineering research and practice has paid little attention to the issues involved and the most relevant background work comes from the discipline of systems engineering. Systems engineering focuses on the development of systems as a whole, as defined by the International Council for Systems Engineering (INCOSE) (http://www.incose.org/practice/whatissystemseng.aspx):

*"Systems Engineering integrates all the disciplines and specialty groups into a team effort forming a structured development process that proceeds from concept to production to operation. Systems Engineering considers both the business and the technical needs of all customers with the goal of providing a quality product that meets the user needs."*

Systems engineering emerged to take an overall system-wide perspective on complex engineered systems involving structures, electrical and mechanical systems. Now almost all systems are 'software-intensive' and the problems of constructing ultra-large scale software systems are being addressed by this community (Sillitto, 2010).

Work from systems engineering that is particularly relevant to this paper is 'system of systems' (SoS) research (Maier, 1998). Maier argues that the distinction between a system of systems and a complex monolithic system is that the elements of a SoS are operationally and managerially independent. He presents a characterization of different types of SoS from directed (systems developed for a particular purpose) to virtual (systems that lack a central management authority or centrally-agreed purpose).

Unfortunately, terminology in this area can be confusing. Our view is that the use of the term 'system' implies that, irrespective of the components, the entity that is created is purposeful (Checkland, 1981) – intentionally designed to serve some organizational purpose or need. This is consistent with the definition of system of systems proposed by the US Department of Defense (DoD, 2008):

"*An SoS is defined as a set or arrangement of systems that results when independent and useful systems are integrated into a larger system that delivers unique capabilities*"

The implication is that an SoS is created by a single organization (e.g. the US Air Force) that integrates internal and external systems to do something that serves some purpose for that specific organization. There is an 'owner' of the whole system who has at least some influence over the constituent systems and who can certainly decide which systems are components of the SoS.

The interacting algos that led to the Flash Crash are owned by different organisations and may be systems of systems in their own right. They serve the different purposes of their owners and they only cooperate because they have to. The owners of the individual organizational systems are often competing and may be mutually hostile. Each system jealously guards its own information and may change without consultation with any other systems.

In Maier's terms, the collection of systems that led to the Flash Crash would be called a 'virtual system of systems'. However, the prefix 'virtual' is not consistent with other common usage of that term e.g. 'virtual machines'.

Rather than using the unintuitive term 'virtual system of systems', we propose an alternative namely 'coalition of systems'. A coalition of systems is a collection of systems that work together, sometimes reluctantly, because it is in their mutual interest to do so. Coalitions of systems are not explicitly designed but come into existence when different systems interact according to agreed protocols. Like political coalitions, there may be hostility between the members and members may enter and leave the coalition according to their interpretation of what is in their best interests.

Coalitions of software systems make software engineering particularly challenging. We can't design dependability into the coalition as there is no overall design authority; nor can we control the behaviour of individual systems. The systems in the coalition may change unpredictably, may be completely replaced and the organizations running these systems may themselves go out of existence. Coalition 'design' involves designing the protocols for communications and each organization using the coalition orchestrate the constituent systems in their own way. However, the designers and managers of each individual system have to consider how to make their systems robust enough to ensure that their organizations are not threatened by failures or any undesirable behaviour elsewhere in the coalition.

By analogy with Rittel and Weber's notion of 'wicked problems' (1973), coalitions of systems can be thought of as 'wicked systems' (Metcalfe, 2004). Wicked problems are impossible to completely understand as they change as we attempt to address the problem. 'Wicked systems', similarly, are constantly changing as they are developed and used and are impossible to understand completely.

## 3. IT System complexity

The complexity of a system stems from the number and type of relationships between the system components and between the system and its environment. If there are a relatively small number of relationships between system components and these change relatively slowly over time then we can develop deterministic models of the system and make predictions of its properties.

However, when there are many dynamic relationships between the elements in a system then we have a complex system. Complex systems are non-deterministic and system characteristics cannot be predicted by analysis of the system constituents. These characteristics emerge when the whole system is put into use and they change over time, depending on how the system is used and the system's external environment.

Dynamic relationships include relationships between system elements and the system's environment that change. For example, a trust relationship is a dynamic relationship. Initially, component A may not trust component B so, after some interchange, A checks that B has performed as expected. Over time, these checks may be reduced in scope as A's trust in B increases. However, some failure in B may then profoundly influence that trust and after failure, even more stringent checks may be introduced.

Complexity that stems from the dynamic relationships between the elements in a system is '*inherent complexity*' – it depends on the number, existence and nature of these relationships. We cannot analyse inherent complexity during system development as it depends on the system's dynamic operating environment. Coalitions of systems whose elements are large software systems will always be inherently complex. The relationships between the elements of the coalition change because they are not independent of the ways that the constituent systems are used and their operating environments. Consequently, the non-functional and, often, the functional behaviour of coalitions of systems is emergent and impossible to completely predict.

However, even when the relationships between system elements are simpler, relatively static and, in principle, understandable, there may be so many elements and relationships that understanding these relationships is practically impossible. This type of complexity is '*epistemic complexity*' – it stems from our lack of knowledge about the system rather than inherent system characteristics (Rushby, 2009). For example, it may be possible in principle to deduce the traceability relationships between requirements and design but, if the appropriate tools are not available, then it may be practically impossible to do so.

If you don't know enough about a system's components and their relationships, you cannot make predictions about it, even if that system does not have dynamic relationships between its elements. Epistemic complexity increases with the size of the system – as we build larger and larger systems, it is inevitable that they will become harder to understand and their behaviour and properties will be harder to predict.

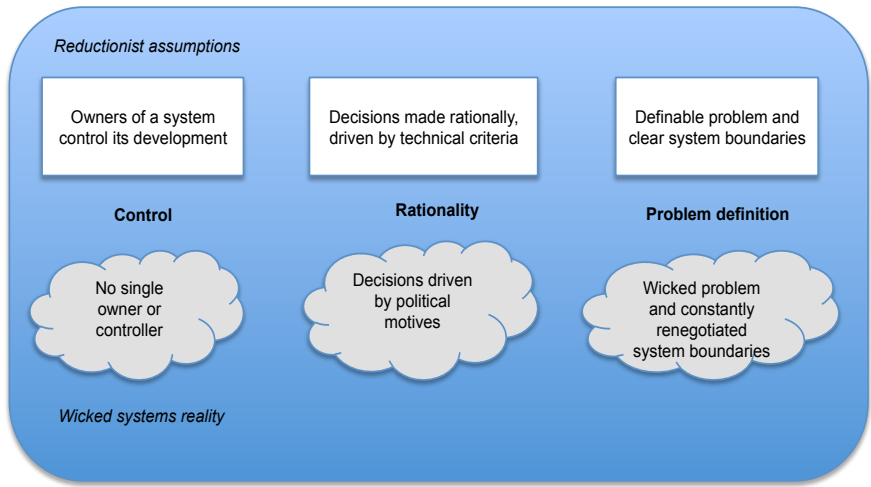

**Figure 1: Reductionist assumptions and wicked systems**

This distinction between inherent and epistemic complexity is important. As we discuss in the following section, we believe that it is the primary reason why we need new approaches to software engineering.

## 4. Reductionism and software engineering

In some respects, software engineering has been incredibly successful. Compared to the systems that were being built in the 1970s and 1980s, modern software is much larger, considerably more reliable and often developed more quickly. Software products deliver astonishing functionality for relatively low prices.

Software engineering has focused on reducing and managing epistemic complexity so, where inherent complexity is relatively low and, critically, where a single organization controls all elements of the system, software engineering is very effective. However, we argue that for coalitions of systems with a high degree of inherent complexity, current software engineering techniques are inadequate.

We see this in the failures that are common in large government-funded projects. The software is delivered late, is more expensive to develop than anticipated and does not meet the needs of its users. An example of such a project was the automation of health records in the UK where the project was abandoned after 10 years of development. Estimates of the costs of this failure range from 5 to 10 billion dollars.

We argue that there is a fundamental reason why current software engineering cannot effectively manage inherent complexity, with the consequence that our software engineering methods are unsuitable for building $21^{st}$ century wicked systems. To understand this, we need to examine the essential 'divide-and conquer' reductionist assumption that is the basis for modern engineering.

Reductionism is a philosophical position that a complex system is nothing but the sum of its parts, and that an account of it can be reduced to accounts of individual constituents. From an engineering perspective, this means that you should design a system so that it is composed of discrete, smaller parts and define interfaces that allow these parts to work together. You then build the system elements and integrate these to create the desired system.

Researchers in software engineering generally adopt this reductionist assumption and their work has either been around finding better ways to decompose problems or systems (e.g. work in software architecture), better ways to create the parts of the system (e.g. object-oriented techniques) or better ways of system integration (e.g. test-first development).

There are three fundamental reductionist assumptions that underlie software engineering methods and techniques as shown in Figure 1:

1. *Owners of a system control its development.* A reductionist perspective takes the view that there is an ultimate controller who has the authority to take decisions about a system and who can therefore 'enforce' decisions on, for example, how components should interact. However, when systems are composed of independently owned and managed elements, there is no such owner or controller and there is no central authority to make or enforce design decisions.

2. *Decisions are made rationally and are driven by technical criteria.* In fact, decision making in organizations is profoundly influenced by political considerations where actors in the organization strive to maintain or improve their current position or avoid losing face. Technical considerations are rarely the most significant factor in large system decision making.

3. *There is a definable problem and clear system boundaries.* The nature of 'wicked problems' is that the 'problem' is constantly changing depending on the perception of stakeholders and external events. The system boundaries are influenced by both these changes and the status and perspectives of stakeholders. As these change, the boundaries are redefined.

For wicked systems, these assumptions are never true and many software project 'failures', where software is delivered late and over-budget, are a consequence of adherence to this reductionist view of the world. To help us address inherent complexity, software engineering must evolve to look outwards and to embrace the other systems, people and organizations that make up the software systems' environment. We need to represent, analyze, model and simulate potential operational environments for coalitions of systems to help us understand and manage, so far as is it is possible to do so, the complex relationships in the coalition.

## 5. Challenges for research and education

There are initiatives in the US and in Europe that are starting to address the problem of engineering large, complex, coalitions of systems. In this US, the influential SEI report (Northrop *et. al* 2006) on Ultra-Large Scale Systems (ULSS) has triggered research at the SEI and has led to the creation of ULSSIS, a research consortium involving Virginia, Michigan State, Vanderbilt Universities and UCSD. In the UK, the Large-Scale Complex IT systems (LSCITS) initiative is addressing problems of both inherent and epistemic complexity in large-scale IT systems and Hillary Sillitto from Thales is considering design principles for ULSS (2010).

Northrop *et al.* have made the point that we need to go beyond incremental improvements to current methods and have identified research areas that are important for ULSS namely human interaction, computational emergence, design, computational engineering, adaptive system infrastructure, adaptable and predictable system quality and policy, acquisition and management. Their report suggests that it is essential to deploy expertise from a range of disciplines to address these problems.

We are in complete agreement that the research required is inter-disciplinary and that incremental improvements in existing techniques will be insufficient to address future software engineering challenges. However, 'breakthroughs' in engineering research are uncommon and take many years to be exploited. Across the world, we are now engineering large complex software systems and so there is also a need for a more immediate, perhaps more incremental, research, driven by the practical problems of complex IT systems engineering.

Part of this will involve developing and extending current software engineering methods. Epistemic complexity will continue to increase as software systems get larger and larger and we need software engineering techniques, such as formal analysis and modelling, to help deal with this. But most of the work has to focus on the challenges posed by complexity.

## 5.1 A research agenda for software engineering

The engineering of coalitions of systems involves the engineering of individual systems so that they can work effectively in a coalition and the orchestration and configuration of a coalition of systems to meet some organizational needs. Based on the ideas in the ULSS book and our own experience in the UK LSCITS initiative, we have identified the 'top-10' problems that define a research agenda for future software systems engineering.

1. How can we model and simulate the interactions between independent systems?

   To help us understand and manage coalitions of systems we need dynamic models that are updated in real-time with information from the actual system. We need these models to help us make rapid 'what-if' assessments of the consequences of system change options. This will require new performance and failure modelling techniques where the models can adapt automatically from system monitoring data. Of course, we are not suggesting that simulations can be complete or can predict all possible problems. However, other engineering disciplines have benefited enormously from simulation and we believe that comparable benefits might be achieved for software engineering.

2. How can we monitor coalitions of systems and what are the warning signs of problems?

   In the run-up to the Flash Crash there were no indicators that might have indicated that the system was tending towards an unstable state. To help avoid the transition to an unstable system state, we need to know what are the indicators that provide information about the state of the coalition of systems, how these indicators may be used to provide both early warnings of system problems and, if necessary, switch to safe-mode operating conditions that will stop damage occurring. To make effective use of this data, we need visualization techniques that reveal the subtleties of coalition operation and interactions to operators and users.

3. How can systems be designed to recover from failure?

   A fundamental principle of software engineering is that we should build systems so that they do not 'fail'. This has led to the development of methods and tools based on fault avoidance, fault detection and fault tolerance.

   However, as we construct coalitions of systems with independently-managed elements and negotiated requirements, it is increasingly impractical to avoid 'failure'. Indeed, what seems to be a 'failure' for some users may not affect some others. Because some failures are ambiguous, automated systems cannot cope on their own. Human operators have to use information from the system and intervene to recover from the failure and restore the system. This means that we need to understand the socio-technical processes of failure recovery, the support that these operators need and how to design coalition members to be 'good citizens' and to support failure recovery.

4. How can we integrate socio-technical factors into systems and software engineering methods?

   Software and systems engineering methods have been created to support the development of technical systems and, by and large, consider human, social and organisational issues to be outside the system boundary. However, these non-technical factors significantly affect the development, integration and operation of coalitions of systems. There is a considerable body of work on socio-technical systems but this has not been 'industrialised' and made accessible to practitioners. A recent paper (Baxter and Sommerville, 2010) surveys this work and proposes a route to industrial use of socio-technical methods. However, much more research and experience is required before socio-technical analyses can be routinely used for complex systems engineering.

5. To what extent can coalitions of systems be self-managing?

    The coalitions of systems that will be created are complex and dynamic and it will be difficult to keep track of system operation and respond in a timely way to the monitoring and health measurement information that is provided. We need research into self-management so that systems can detect changes in both their own operation and in their operational environment and dynamically reconfigure themselves to cope with these changes. The danger is that reconfiguration will create further problems so a key requirement is for these techniques to operate in a safe, predictable and auditable way and to ensure that self-management does not conflict with 'design for recovery'.

6. How can we manage complex, dynamically changing system configurations?

    Coalitions of systems will be constructed by orchestration and configuration and the desired system configurations will change dynamically in response to load, indicators of the system health, unavailability of components and system health warnings. We need ways of supporting construction by configuration, managing configuration changes and recording changes (including automated changes from the self-management system) in real-time so that we have an audit trail recording what the configuration of the coalition was at any point in time.

7. How can we support the agile engineering of coalitions of systems?

    The business environment changes incredibly quickly in response to economic circumstances, competition and business reorganization. The coalitions of systems that we create will have to change rapidly to reflect new business needs. A model of system change that relies on lengthy processes of requirements analysis and approval simply will not work.

    Agile methods of programming have been successful for small to medium sized systems where the dominant activity is system development. For large and complex systems, development processes are often dominated by coordination activities involving multiple stakeholders and engineers who are not co-located. How can we evolve agile approaches that are effective for 'systems development in the small' to support multi-organization, global software development?

8. How should coalitions of systems be regulated and certified?

    Many coalitions of systems will be critical systems whose failure could threaten individuals, organizations and economies. They may have to be certified by a regulator who will check that, as far as possible, the systems will not pose a threat to their operators or the wider systems' environment. But certification is increasingly expensive. For some safety-critical systems the cost of certification may exceed the costs of development. These costs will continue to rise as systems become larger and more complex.

    Although certification as currently practised is almost certainly impossible for coalitions of systems, there is an urgent need for research that allows for incremental and evolutionary certification so that our ability to deploy critical complex systems is not curtailed by the certification requirements. This is a social as well as a technical issue – our societies have to decide what level of certification is socially and legally acceptable.

9. How can we do 'probabilistic verification' of systems?

    Our current techniques of system testing and more formal analysis are based on the assumption that the system has a definitive specification and that behaviour which deviates from that specification can be recognized. Coalitions of systems will have no such specification nor will system behaviour be guaranteed to be deterministic. The key

verification issue will not be 'is the system correct' but 'what is the probability that it satisfies essential properties, such as safety, that take into account its probabilistic, real-time and non-deterministic behaviour' (Kwaitowska *et al.,* 2009; Ge *at al.,* 2010).

10. How should shared knowledge in a coalition of systems be represented?

    We assume that the systems in a coalition will interact through service interfaces so there will not be any over-arching controller in the system. Information will be encoded in a standards-based representation. The key problem will not therefore be a problem of compatibility – it will be a problem of understanding what the information that systems exchange actually means.

    Currently, we address this problem on a system by system basis with negotiations taking place between system owners to clarify what shared information means. However, if we allow for dynamic coalitions with systems entering and leaving the coalition, this is no longer a practical approach. The key issue is developing a means of sharing the meaning of information – perhaps using ontologies as proposed in the work on the semantic web (Antoniou and Van Harmelan, 2008).

A major problem that researchers face is a lack of knowledge of what currently happens in real systems. High-profile failures, such as the Flash Crash, lead to enquiries but we need more information about the practical difficulties faced by developers and operators of coalitions of systems and how they cope with problems that arise. New ideas, tools and methods, need to be supported by long-term empirical studies of these systems and their development processes to provide a solid information base to inform research and innovation.

The LSCITS project (Cliff *et al.* 2010; Cliff and Northrop, 2011) is tackling some of these issues. We are working with partners from the computer industry, financial services and healthcare to develop an understanding of the fundamental systems engineering problems that they face. We have a long-term engagement with the UK body that manages national healthcare data who need to create coalitions of systems for external access and analysis of the vast amounts of data involved.

We are developing practical techniques of socio-technical systems engineering (Baxter and Sommerville, 2010) and are exploring the problems of designing for failure (Sommerville, 2008). We have developed practical and predictable techniques for autonomic system management (Calinescu and Kwiatkowska, 2009, Calinescu *et al*, 2011) and are investigating the scaling-up of agile methods (Paige et al., 2008). We are exploring possibilities for incremental system certification (Ge *et al.,* 2010) and are working on the development of techniques for system simulation and modelling.

## 5.2 Education

To help us address the practical issues of creating, managing and operating wicked systems, we need engineers who are equipped with knowledge and understanding of the challenges posed by these systems and with techniques that lie outside a 'normal' software or systems engineering education.

In the UK, we are providing graduate student education with a new kind of doctoral degree, comparable in standard to a PhD. Our students get an Engineering Doctorate (EngD) in Large-scale complex IT systems, (University of York, 2009) where the key differences between an EngD and a PhD are:

1. Students have to work on an industrial problem and spend a significant period of time working in industry on that problem. Universities simply cannot replicate the complexity of modern software-intensive systems and few faculty have experience and understanding of these systems.

2. Students take a range of courses that focus on complexity and systems engineering such as systems engineering for LSCITS, socio-technical systems, high-integrity systems engineering, empirical methods and technology innovation.
3. Students don't have to produce a conventional 'thesis' – a book on a single topic but can produce a portfolio of work around their selected area. This is a better reflection of the reality of work in industry and makes it easier for the topic to evolve as systems change and new research emerges.

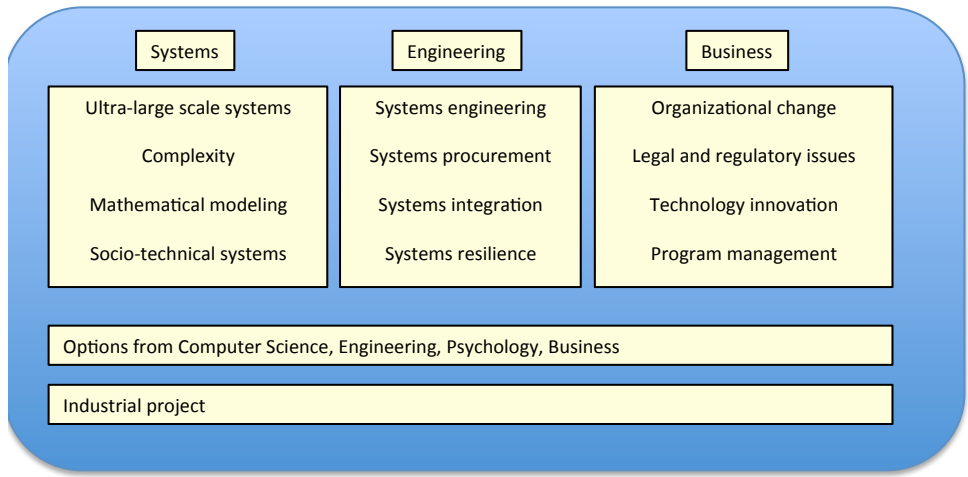

**Figure 2: Outline structure for Masters course in Large-Scale Complex IT Systems**

However, graduating a few advanced doctoral students is simply not enough. Universities and industry now need to work together to create Masters courses that educate complex systems engineers for the 21$^{st}$ century. Figure 2 sets out our thoughts on what might be covered in such courses. We understand that a comparable course in ULSS is being developed at Queens University in Canada but, at the time of writing, no details on this are available.

Masters courses in this area have to be multidisciplinary, bringing together engineering and business disciplines. It is not only the knowledge that these disciplines bring that is important. It is also critical that students are sensitised to the perspectives of different disciplines and so move out of the silo of single discipline thinking.

## 6. Conclusions

Since the advent of widespread networking in the 1990s, our societies have grown increasingly dependent on complex software-intensive systems. Serious failures of these systems will have profound social and economic consequences. As Sillitto (2010) says, we are building these systems without an understanding of how to analyze their behaviour and without appropriate engineering principles to support their construction.

We have argued here that there are fundamental reasons why existing approaches cannot be 'scaled-up' to create coalitions of systems and that incremental improvements to today's methods are not enough to cope with complexity. Put bluntly, existing software engineering is not good enough. We need to think differently to address the urgent and growing need for new engineering approaches that can help us construct complex systems that we can trust.

*Sidebar: Socio-technical systems*

Engineers are primarily concerned with building technical systems with hardware and software components. They assume that the system requirements reflect the organizational needs for integration with other systems, compliance and business processes. Yet, when we consider systems in use, these are not simply technical systems but 'socio-technical systems'. To reflect the fact that these are evolving and interacting communities that include technical, human and organisational elements, these are sometimes called 'socio-technical ecosystems'. However, the term 'socio-technical systems' is one that is more commonly used.

Socio-technical systems are organizational systems that include people and processes as well as technological systems. The process definitions set out the intentions of the system designers as to how the system should be used but, in reality, the people in the system interpret and adapt these in a range of different ways depending on their education, experience and culture. Individual and group behaviours also depend on organizational rules and regulations as well as 'organizational culture' – 'the way we do things around here'.

An over-simplification that has hindered software engineering for many years is that it is possible to consider technical software-intensive systems that are intended to support work in an organization in isolation. The so-called 'system requirements' are the interface between the technical system and the wider socio-technical system yet we know that requirements are inevitably incomplete, incorrect and out of date.

Coalitions of systems cannot operate on this basis. Rather, we must recognize that these are rich socio-technical systems and by taking advantage of the abilities and inventiveness of people we can create more effective and more resilient systems.

## Acknowledgments

We would like to thank our colleagues Gordon Baxter and John Rooksby of St Andrews University and Hillary Sillitto of the Thales Group for their constructive comments on drafts of this paper. We would also like to thank the reviewers of earlier versions of this paper for their helpful and constructive comments.

The work here was partially funded by the UK Engineering and Physical Science Research Council, grant number EP/F001096/1.

## Authors

Ian Sommerville is a Professor in the School of Computer Science, St Andrews University, Scotland.

Dave Cliff is a Professor in the Department of Computer Science, Bristol University, England.

Radu Calinescu is a Lecturer in the Departent of Computer Science, Aston University, England

Justin Keen is a Professor in the School of Health Informatics, Leeds University, England

Tim Kelly is a Senior Lecturer in in the Department of Computer Science, York University, England.

Marta Kwaitkowska is a Professor in the Department of Computer Science, Oxford University England.

John McDermid is a Professor in the Department of Computer Science, York University, England.

Richard Paige is a Professor in the Department of Computer Science, York University, England.